\documentclass[graybox]{svmult}

\usepackage{amsmath}

\usepackage{mathptmx}       
\usepackage{helvet}
\usepackage{courier}        
\usepackage{type1cm}        
\usepackage{makeidx}         
\usepackage{graphicx}        
\usepackage{multicol}        
\usepackage[bottom]{footmisc}

\makeindex             

\usepackage{natbib}
\def\apj{ApJ}
\def\mnras{MNRAS}
\def\aj{AJ}
\def\apjl{ApJ Lett}

\begin{document}
\bibliographystyle{spbasic}

\title*{Tidal Debris as a Dark Matter Probe}
\author{Kathryn V. Johnston and Raymond G. Carlberg}
\institute{Columbia University; University of Toronto}
%
%

\motto{Chapter~7 from the volume ``Tidal Streams in the Local Group and Beyond: Observations and Implications''; ed. Newberg, H.~J., \& Carlin, J.~L.\ 2016, Springer International Publishing, Astrophysics and Space Science Library, 420\\ ISBN 978-3-319-19335-9; DOI 10.1007/978-3-319-19336-6\_1}
\maketitle

\abstract{Tidal debris streams from galaxy satellites can provide insight into the dark matter distribution in halos.  This is because we have more information about stars in a debris structure than about a purely random population of stars: we know that in the past they were all bound to the same dwarf galaxy; and we know that they form a dynamically cold population moving on similar orbits.  They also probe a different region of the matter distribution in a galaxy than many other methods of mass determination, as their orbits take them far beyond the typical extent of those for the bulk of stars. 
Although conclusive results from this information have yet to be obtained, significant progress has been made in developing the methodologies for determining both the global mass distribution of the Milky Way's dark matter halo and the amount of dark matter substructure within it.  Methods for measuring the halo shape are divided into ``predictive methods," which predict the tidal debris properties from the progenitor satellite's mass and orbit, given an assumed parent galaxy mass distribution; and ``fundamental methods," which exploit properties fundamental to the nature of tidal debris as global potential constraints.  Methods for quantifying the prevalence of dark matter subhalos within halos through the analysis of the gaps left in tidal streams after these substructures pass through them are reviewed.}

\section{Introduction}
\label{intro.sec}

Understanding how matter is distributed in galaxies is a fundamental problem in astronomy.
In particular, cosmological simulations of structure formation within the standard $\Lambda$CDM model of the Universe suggest that the stars we see collected together as  galaxies are surrounded by much more massive and extended dark matter halos\index{dark matter halos}.
The simulations outline expectations for the average properties of halos, including their characteristic density profiles \citep{nfw97}\index{Navarro, Frenk, \& White (NFW) profile} and triaxial shapes and orientations as a function of radius \citep{jing02}.
They also suggest that each Milky-Way-sized dark matter halo encompasses a swarm of 
satellite subhalos\index{dark matter subhalos!satellite} in numbers far greater than the number of observed satellite dwarf galaxies in the Milky Way halo
 \citep[traditionally referred to as the ``missing satellites''\index{missing satellites} problem, see][]{moore99,klypin99}.
While it is possible that the observed dwarf galaxies account for all of the largest of these satellite subhalos \citep[though this is far from clear, see][]{boylan-kolchin11}, a multitude of smaller halos (masses of order $10^7 M_\odot$ or less) are still predicted to exist for which observed counterparts have not been identified (see Section \ref{sec:local} for a more complete discussion).

The expectations for the structure of and substructure within dark matter halos from the cosmological simulations are in general hard to test with great accuracy since we currently observe dark matter only by its gravitational effects on stars, and the dark matter extends well beyond where the majority of the stars in any galaxy lie.
Tracer populations such as planetary nebulae and globular clusters have been used to estimate global masses beyond the visible components of galaxies \citep{cote03}, while satellite systems are thought to provide some accounting of the substructure --- though it remains far from clear how complete and unbiased that accounting is \citep{tollerud08}.
Some progress on the projected shapes, total masses and largest substructures within dark matter halos has come from gravitational lensing \citep{vegetti10}.
This chapter outlines why tidal debris is considered a  promising and sensitive probe of dark matter halos as well as the subhalos they are expected to contain.
In particular, this chapter concentrates on debris around our own Milky Way galaxy since this is the one place in the Universe where we might hope to measure the three-dimensional structure of a dark matter halo, as well as be sensitive to the proposed multitude of lower mass subhalos that may not contain gas or stars.

Historically, the knowledge that stars in disk systems are moving on near-circular orbits has been exploited to sensitively measure their mass distributions --- indeed, the first systematic studies of the rotation curves of galaxies promoted the idea that galaxies were dominated by dark matter rather than baryonic matter \citep{rubin70}.  Although the nature of the dark matter particles themselves has yet to be determined, the idea that the majority of the mass in the Universe is composed of particles that are so far not observed because they do not interact with light is
generally accepted in the field.

Stars in tidal debris structures make excellent probes of the matter distribution around galaxies for analogous reasons to disk stars. 
Like disk stars, we know more about their orbits than we would know for a purely random population --- we know that the stars in tidal streams were once all part of the same parent satellite galaxy, and consequently have a small range of orbital properties about the progenitor satellite's orbit.
Moreover, to be detectable in a photometric study, these debris structures must lie well outside the bulk of the stars in the parent galaxy and hence typically probe a very different region of the dark matter halo than the bulk of the stars in a galaxy.

Images of {\it streams} of debris in particular suggest that stars lie close to a single orbit, and this gives some simple insight into why they are such sensitive potential probes.
If the stars were actually on {\it exactly} the same orbit and you could measure the positions $\vec{x}$ and velocities $\vec{v}$ of each, then the potential (up to the unknown, constant energy $E$ of the orbit) would simply be $\Phi(\vec{x})=E-v^2/2$.
A more detailed discussion of methods for measuring the global potential is presented in \S \ref{sec:global}.

The dynamical {\it coldness} of streams, as determined from the small velocity dispersions and narrow spatial cross sections observed for many streams, also provides simple insight into their use as substructure probes. This low temperature means that discontinuities in the debris on much smaller scales than the apparent orbital path can arise due to asymmetries in the potential that are also on much smaller scale than the global potential. 
This leads to the exciting possibility of using debris to detect substructures within the potential, such as those arising from the thousands of dark matter subhalos that are expected to be orbiting within the main halo.  Direct evidence of a large number of dark subhalos would {\it solve} the missing satellite problem.
The characteristic signatures that substructure may leave in streams are discussed in \S \ref{sec:local}.

\begin{figure}[!t]
\begin{center}
\includegraphics[width=0.8\textwidth]{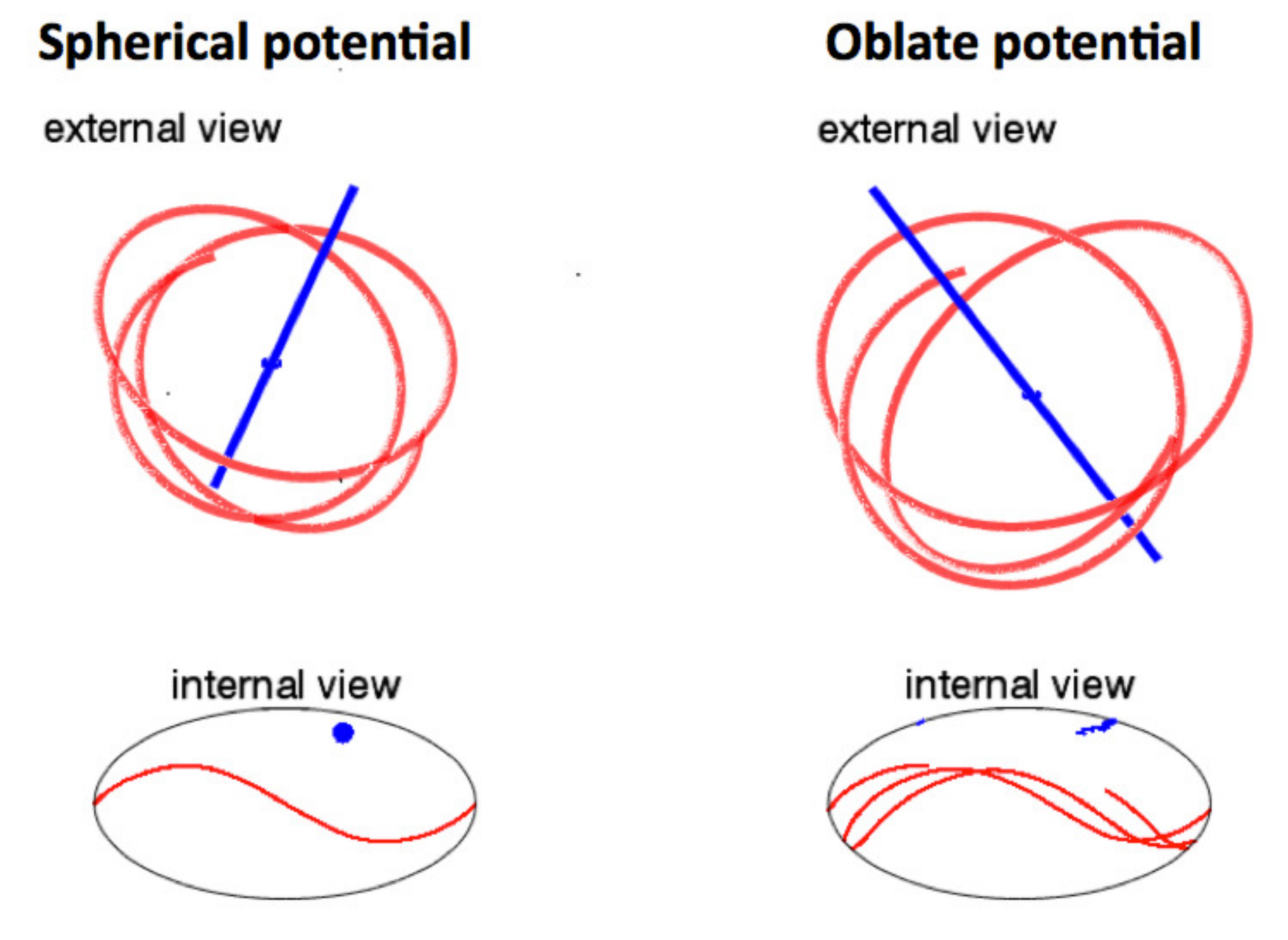}
\end{center}
\caption{The left and right panels show (in red) external and internal 
views of orbital paths in spherical and oblate potentials respectively.  External views are as seen from outside the galaxy; the center of the galaxy is shown by a small blue dot.  Internal views are Aitoff projection all-sky maps, as seen from the galaxy center. The direction of the orbital pole is shown by the blue line in external views and blue dots (in the oblate case a moving dot) in the internal views.  The position of the pole changes over time for the oblate case due to the precession of the orbital plane, but the position of the pole does not change for the spherical case.}
\label{fig:orb_plane}
\end{figure}

\begin{figure}[!t]
\begin{center}
\includegraphics[scale=0.7]{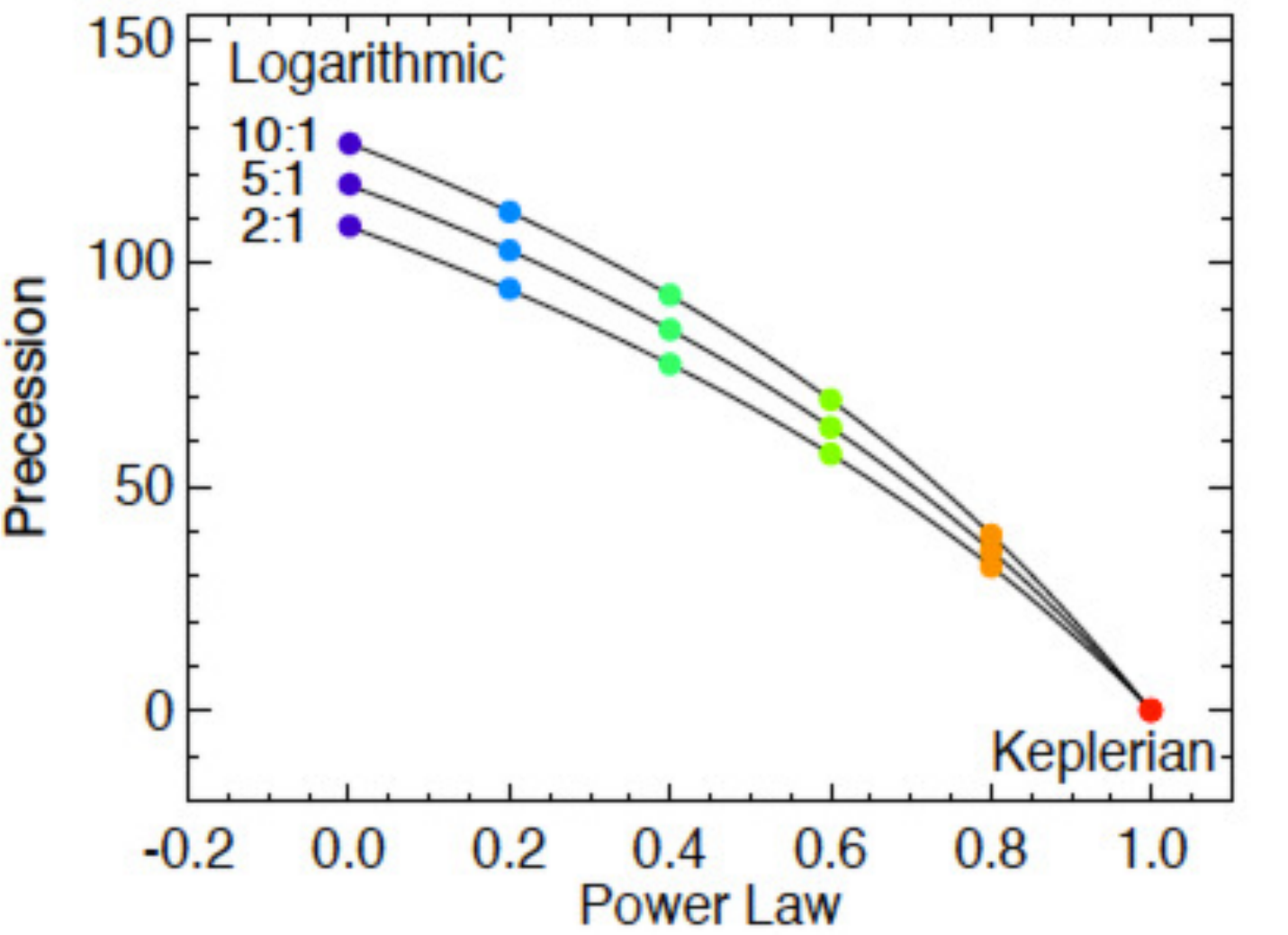}
\end{center}
\caption{Potential profile dependence of the precession of turning points. The plot shows the angle between successive apocenters for potentials varying from logarithmic to Keplerian. The colors are for orbits of different eccentricity. Reproduced from \citet{belokurov14}.}
\label{fig:orb_apo}
\end{figure}

\section{Using tidal debris to probe the global potential}
\label{sec:global}


As indicated in the introduction, adopting the simplifying (but ultimately incorrect) assumption that debris {\it exactly} traces a single orbit allows us to develop some intuition for how this data might be used.
We can expand this intuition a little more  by considering properties of orbits in different potentials. For example:
\begin{enumerate}
\item Orbits in a {\it spherical} potential are purely planar, while in {\it oblate} potentials, the orbital plane itself precesses with time. Figure \ref{fig:orb_plane} illustrates this by showing the orbital path (in red) and orbital pole (i.e., direction of angular momentum, in blue) in 3-D (top panels) and projected onto the plane of the sky (as viewed from the center of the parent galaxy --- bottom panels) for the same orbit in spherical (left panels) and oblate (right panels) potentials.   In the spherical case, the orbit aligns with a single plane on the sky and the pole appears as a single point. In the oblate case, the precession of the orbital plane is apparent as the orbital path not aligning with a single great circle and the direction of the pole evolving in time.
\item Within the orbital plane, the {\it precession of the angular position of apocenters} reflects the radial density profile of the parent potential. Figure \ref{fig:orb_apo} \citep[from][]{belokurov14} shows this precession angle calculated for a variety of potentials and orbital eccentricities, with zero precession corresponding to the point-mass, Keplerian potential in which orbits are ellipses and the apocenters are coincident.
\item The {\it eccentricity} (or ratio of apocenter to pericenter) of the orbit for a given pericentric velocity is  set by the overall depth of the potential. 
\end{enumerate}

We are now in a position to measure the attributes listed above for observed streams.
For example, using debris from the (ongoing) destruction of the Sagittarius (Sgr) dwarf spheroidal (see Chapter~2 of this volume), we know the precession of the orbital plane \citep{majewski03}, the trends of velocity along the stream \citep{majewski04}, as well as the angular separation between successive apocenters \citep{belokurov14}. 
It is this level of detailed data that has enabled the first attempt to reconstruct the 3-dimensional shape and orientation of our Galactic dark matter halo \citep{lm10a}, or indeed of any galaxy in the Universe.
However, significant controversy remains over what this reconstruction means, and this is inspiring a thorough and rigorous examination of how debris can robustly be used to measure potentials.

A multitude of approaches to exploiting our rich debris data sets to measure potentials have been proposed. 
They can be very broadly divided into two categories:
\begin{description}
\item{\it Predictive models} predict the exact six-dimensional phase-space position and density of tidal debris from the progenitor's properties, orbit and parent galaxy potential.
It is trivial to  project the predictions of these models from phase-space to observables and hence to perform the comparison of data and model in coordinates where error distributions are well-understood.
Hence these methods are the easiest for which to account for the effect of errors (or even missing dimensions of information) on any parameter estimates.
However, they are also the most sensitive to biases that may result from incorrect assumptions in the model --- in particular in how stars are distributed in the satellite.
For example, a rotating satellite can produce tidal streams with centroids systematically offset from those produced by a non-rotating satellite \citep{penarrubia10}.

\item{\it Fundamental methods} exploit some more general intrinsic property that debris obeys rather than matching the full density distribution in phase-space.
Typically, these methods require few (if any) assumptions about the satellite's internal structure and may be less prone to such biases than predictive models. 
On the other hand, they always involve a non-trivial transformation from observed co-ordinates to perform a comparison with model predictions, so correcting for systematic and random biases on parameter estimates due to errors in observables is difficult, if not impossible.
\end{description}
Methods in both categories are reviewed in more detail below.

\subsection{Predictive models}

The most obvious example of a predictive model is an N-body simulation in which a ball of particles is allowed to evolve subject to its own gravity and the influence of external forces from the parent galaxy in which it is orbiting.
These models naturally include the physics of tidal stripping and are hence expected to generate the most physically realistic  debris distributions for a satellite of a given mass and internal stellar distribution, orbiting in a given potential.
Comparison of such models with data have been used most extensively in the case of the Sagittarius dwarf galaxy to measure the radial profile, depth and even triaxial shape of the Milky Way's potential  \citep{helmi04,law05,lm10a}.
Nevertheless, the results for Sgr remain perhaps the most controversial, as the size and morphology of Sgr's streams suggest they might have been affected by additional physical effects beyond those incorporated in the current models (such as internal properties,  orbital evolution due to a much larger primordial mass, or encounters with another satellite; see Chapter~2 of this volume for a more complete discussion).

The downside of a purely N-body approach is the computational cost of the models. 
\citet{lm10a} performed perhaps the most complete exploration of parameter space in any N-body study by identifying a modest region of parameter space using simple test-particle methods and then exploring this region with a grid of models to find a ``best'' fit by brute force.
The ever-improving quality and quantity of data demands a more sophisticated, automated approach to exploring possible models, which in turn motivates exploration of how more approximate but computationally cheaper methods might be utilized.

The cheapest and most trivial example of a predictive model is to assume that a stream traces a single orbit (e.g., \citealt{jlm05,koposov10,deg13}). Note that \citet{binney08} describes how this principle could be exploited even in the case of missing data dimensions. However, this approach cannot be used as anything more than broadly indicative of the behavior of the potential as studies have repeatedly revealed that debris occupies a range of orbits with properties systematically offset from the satellite \citep{johnston98,helmi99a,johnston01,eyre11,bovy14} and demonstrated that this offset leads to systematic biases in potential parameters \citep{sanders13a,lux13}. Recent work in this area has taken the approach of instead using test-particle explorations to outline which streams (or which combination of streams) might be the most informative in measuring the Galactic potential {\it assuming} that these biases can be corrected for \citep{lux13,deg14}. 

A number of methods have been proposed that move beyond the single-orbit approximation, without resorting to N-body simulations. All rely on our knowledge of scales in tidal debris gleaned from our understanding of the physics of tidal limitation and disruption as well as simulations of this process, as outlined in section 6.3.2 and 6.3.3.1 of Chapter~6. 
For example, once lost from a satellite, the evolution of debris can be reasonably represented by test particle orbits \citep[though care has to be taken in setting up the initial conditions for this unbound debris, see][]{gibbons14}, so an approach adopted by several authors is to follow many orbits with properties offset from the satellite's own over scales observed in full N-body simulations  \citep{varghese11,kuepper12,gibbons14}.
In these methods, the satellite's own orbit is followed with additional, offset debris orbits being initialized as the satellite loses mass, and subsequently integrated.
The offset orbits represent the stream properties.

Section 6.3.3.2 of Chapter 6 outlines other predictive models for streams that also start from the  orbit of the satellite, but do not rely on additional particles to represent the debris.
Instead, these methods  calculate the phase-space structure  of stream populations offset from the satellite path given the orbital-phase and time since the material was lost using analytic approximations \citep{johnston98,bovy14,sanders14b}. 
These methods can also be used to search through trial potentials to find a good fit to stream data \citep{johnston99b,sanders14b}.
However, they are limited by the extent to which the adopted analytic approximations apply, or are at least accurate enough for the purposes of recovering the potential. 
For example the model of \citet{johnston98} is based on a description of debris scalings in energy and angular momenta and hence is only strictly applicable to purely spherical potentials. 
The methods proposed by \citet{bovy14} and \citet{sanders14b} are instead formulated in action-angle space and hence provide elegant descriptions for a much wider range of some non-spherical potentials.\footnote{
As discussed in Chapter 6, while there are a very limited number of potentials for which exact analytic actions are known, there has been recent progress in various approximate techniques for finding actions more generally \citep{sanders12,bovy14,sanders15}.
These advances are promising, but the extent of their effectiveness for the purposes of generating accurate models of streams in realistic triaxial potentials has yet to be fully assessed.}

\subsection{Fundamental methods}

We use the term ``fundamental methods'' to refer to 
potential-measuring algorithms that do not generate full models of the phase-space distribution of debris structures, but rather exploit some basic principle that debris must obey. 

For example, \citet{helmi99a} were the first to point out that the accreted nature of stars within a random population in the halo might be uncovered by looking at their orbital properties: stars accreted from a single object would be clustered around the original orbit of the parent satellite.
\citet{helmi99a} used this idea to search for debris in energy and angular momenta in the Solar Neighborhood and \citet{helmi00,gomez10b} went on to explore how these clusters might appear in action-space for a {\it Gaia}-like survey of the halo. 
More recently, \citet{sanderson14} have noted that, since orbital properties (energies and actions) depend on the form of the Galactic potential,  these same ideas could be used to constrain the mass distribution around our Galaxy.
If a significant fraction of the stellar halo is composed from several long-dead satellites then the stars in a random survey  should {\it not} appear random in orbital property-space, but rather clustered.
However, if the orbital properties are calculated in a potential that is {\it not} a good representation of our Galaxy, then the clustering in orbital properties will be less apparent: only in the correct potential is the clustering maximized. 
\citet{penarrubia12} proposed an analogous approach using {\it entropy} as the test statistic to be minimized (equivalent to maximizing the  clustering).

Equation~6.5 in Chapter~6 points to another fundamental property that debris must obey: in action-angle space, the angular offset ($\Delta \theta$)  of debris from the satellite must lie along the same direction as the orbital frequency offset ($\Delta \Omega$) \citep{helmi99a,sanders13b}. 
Since both the angles and the frequencies depend on the form of the potential, this requirement can be used as a potential measure with the correct potential being the one in which the vectors are most closely aligned in the same direction.
\citep[Note that this method implicitly assumes that the debris is distributed isotropically in action space, which is not strictly true --- as pointed out by][]{bovy14}.

Finally, the common origin of debris can be exploited in another way --- if the orbits of stars that are all part of the same debris structure are integrated backwards then their paths should all at some point coincide with the instantaneous phase-space position of the satellite from which they came. Only in the correct potential will this ``recombination'' happen \citep{johnston99a,price-whelan13,price-whelan14}.

In some ways, these methods are very powerful in that they require fewer (if any) assumptions about the properties of the satellite that created the debris. 
In addition, the statistical nature of the approaches of  both \citet{sanderson14} and \citet{penarrubia12}  have the great advantage of not needing to have clear streams already identified in their data sets in order to work. 

\subsection{Summary: status and prospects}

The many papers that have thus far used data on tidal debris to actually measure the properties of the Milky Way's dark matter halo are illustrative both of the potential power of this approach as a well as the extent of the data available (e.g., \citealt{johnston99b,ibata01a,jlm05,koposov10,lm10a,newberg10}).
 However, these works have generally either used the data in very simplistic ways or employed debris models that have not been thoroughly tested.

The field is rapidly maturing with the prospect of much larger and more accurate data sets in the near future motivating the recent development of more sophisticated potential recovery algorithms (outlined above), many of which have been tested with N-body models. 
However, a number of other issues need to be more thoroughly explored:
\begin{itemize}
\item 
Most of the algorithms have only been tested on perfect data and only a couple have attempted to incorporate a rigorous treatment of observational errors \citep[e.g.,][]{koposov10,price-whelan14,gibbons14}.
\item
Tests so far have typically asked how well parameters can be recovered for some assumed  form for the Milky Way's potential. Ideally the Milky Way would be represented in a non-parametric way, allowing more flexibility in the representation of the underlying mass distribution.
For example the mass distribution could be defined by a spatial grid of values or by  the coefficients of a basis function expansion.
\item
Methods that have been formulated in action-space \citep{penarrubia12,sanders13b,bovy14,sanders14b,sanderson14} rely on being able to represent the Milky Way with an integrable potential in which actions can be found.
Recent work has suggested ways of approximately recovering actions for any potential \citep{sanders15,bovy14}.
For example, the true potential can be represented by the sum of a series of integrable potentials with known actions.
However, the level of bias that these approximations introduce into potential-recovery have not yet been explored.
\item 
All the algorithms have been developed and tested only for static, smooth and non-evolving potentials, while we know that the Milky Way has grown and contains many, moving substructures. In a preliminary investigation of these effects, \citet{bonaca14} show that individual stream measurements of the mass of the Milky Way in such an environment can vary by several tens of percent.
\item None of the proposed algorithms have yet investigated the effect of  the internal dynamical distribution in the progenitor satellite, which is known to affect the properties of streams \citep{penarrubia10}.
\end{itemize}

\begin{figure}[!t]
\begin{center}
\includegraphics[scale=0.5]{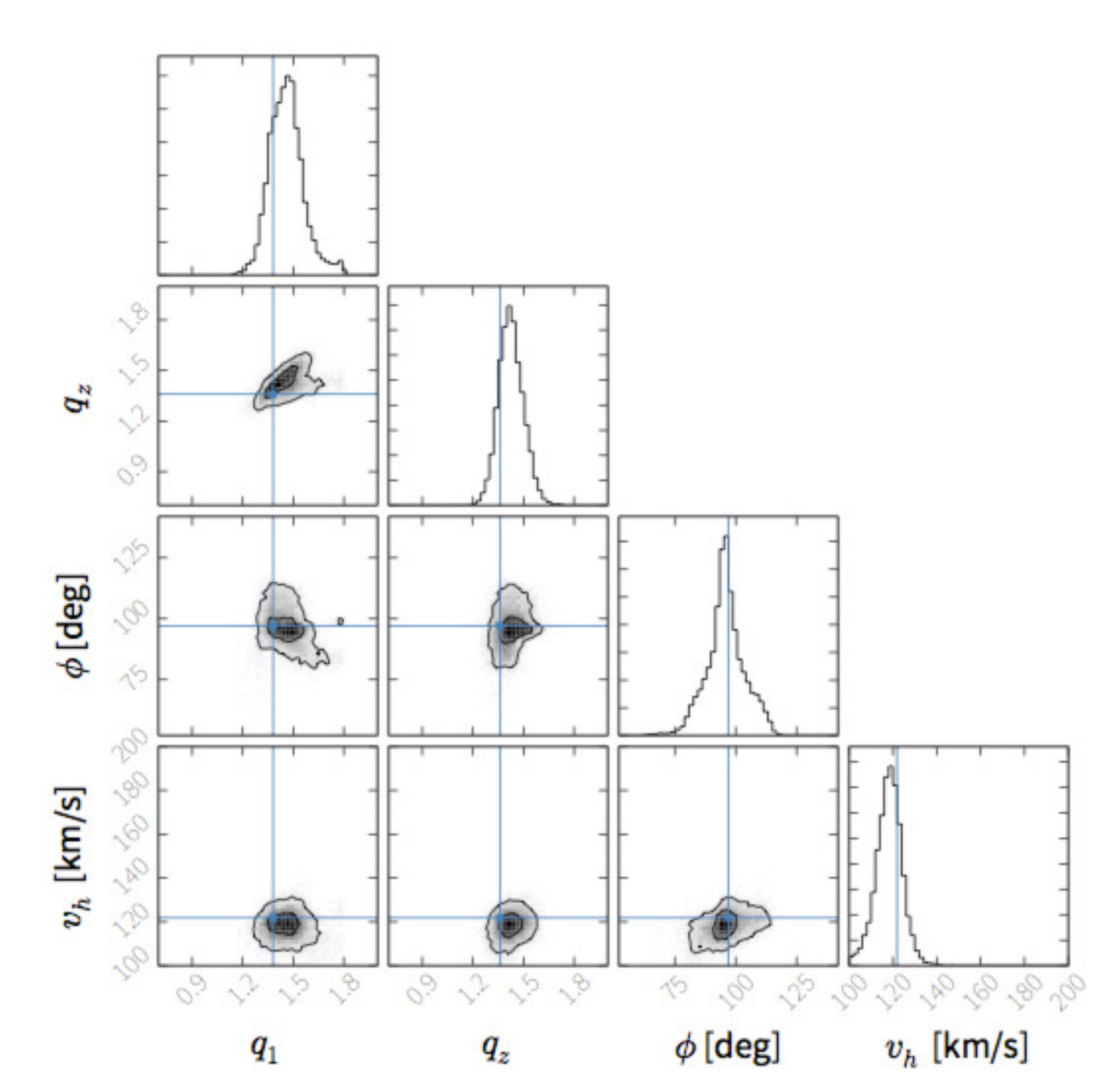}
\end{center}
\caption{Example using the results from the {\tt Rewinder} algorithm to illustrate the power of the tidal tails as potential measures.  A model galaxy containing a tidally disrupted dwarf galaxy was simulated.  Four debris particles were then ``observed'' with errors expected for RR Lyrae stars surveyed with Spitzer \citep[i.e., 2\% distances from the mid-IR period-luminosity relation, see][]{madore12}, Gaia (i.e., very accurate proper motions) and ground-based radial velocity errors of 5 km~s$^{-1}$. In this case, four potential parameters (the velocity scale $v_h$, axis ratios $q_1, q_z$ and orientation $\phi$ of the dark matter halo component in which the simulation was run) were recovered with few percent accuracies on each.  Reproduced from \citet{price-whelan14}.}
\label{fig:rewinder}
\end{figure}

Despite these current limitations,  the promise of this approach provides ample incentive for further investigation.
In particular, as noted above, while gravitational lensing studies are sensitive to the projected shapes of dark matter halos, the Milky Way is the one place in the Universe where we can look at the shape and orientation of a dark matter halo in three dimensions.
As an illustration of this power, Figure \ref{fig:rewinder} shows some results of  tests of the {\tt Rewinder} algorithm \citep[reproduced from][]{price-whelan14} applied to synthetic observations of just four particles drawn from an N-body simulation of satellite disruption. The observational errors were accounted for using a Bayesian approach during the recovery.
The tests show that in the idealized case, where the form of the smooth and static potential is known, few percent errors on potential parameters are possible using even a very small sample with near-future data sets.
For comparison, current estimates for the  mass of the Milky Way differ by more than a factor of two \citep[e.g.,][]{barber14}.
Since real sample sizes will be orders of magnitude larger than those used in the idealized experiment, the results suggest that there is ample room to introduce more flexible (and hence complex) and even time-dependent potentials that will provide a better representation of the true Milky Way mass distribution.

\section{Using tidal debris to probe dark matter substructure}
\label{sec:local}

\subsection{Cosmological Context}

As discussed in the Introduction to this Chapter, standard $\Lambda$CDM models of the Universe  predict an order of
magnitude more dark matter subhalos within the halos of typical
galaxies than the number of known satellite galaxies orbiting the
Milky Way \citep{klypin99,moore99}
 This discrepancy can partially be
explained by accounting for the incomplete sky coverage of SDSS
and the distance-dependent limit on this survey's sensitivity to
low-surface brightness objects \citep{tollerud08,koposov09}.  
Indeed, models which take this into account and 
consider diffuse, (i.e., undetectable) satellite galaxies can reconcile the number
counts for subhalos \citep{bullock10}. However, when they impose the
suppression of stellar populations in low mass subhalos (which have
masses below $5 \times10^8~M_\odot$) the number of undetectable galaxies
significantly declines and the
prediction of numerous purely dark matter subhalos less massive than
$5 \times 10^8~M_\odot$ remains. 
Proof of the existence (or lack) of these ``missing satellites'' 
could provide an important constraint on the nature of dark
matter, which sets the minimum scale for the formation of dark matter
subhalos.

 In much the same way that there is predicted to be a {\it spectrum} of dark matter subhalos in orbits about the Milky Way, we know there is a {\it spectrum} of tidal debris structures; 
the dominant (more extended and hotter) structures (e.g., Sgr and the Orphan Stream) arise from the  infall of the larger subhalos (i.e., the ones that contain stars) while the thinner and colder streams typically come from globular clusters.
All will be disturbed by subhalo-induced fluctuations in the Milky Way's potential \citep[as first investigated by][]{ibata02,JSH:02,SGV:08}.

 The key question is which streams will be most sensitive to the ``missing satellites'' --- in particular, the multitude of low mass ($M < 10^7 M_\odot$) subhalos that are predicted but whose existence has never been definitively proved.
The more dominant streams have much larger cross sections and thus will encounter such subhalos more frequently, but they are also much thicker and hotter, making the effect of such individual encounters less apparent. 
Hence, to address this question, both the frequency of encounters of different mass subhalos as well as the size of the effect of those subhalos compared to the stream's own distribution must be accounted for \citep[see][for explicit calculations]{yoon11}.
For the spectrum of subhalo masses predicted by $\Lambda$CDM, it has been found that hotter stellar streams, such as Sgr, are large enough to hide the signatures of the many encounters it suffers with smaller subhalos, though the (known) subhalos containing visible satellites could have an observable effect \citep{JSH:02}.
Thinner streams, such as Pal 5 and GD-1, should contain significant fluctuations in density and velocity at degree and sub-degree scales due to dozens of direct encounters with subhalos in the mass range $10^5$ -- $10^7M_\odot$ over their lifetimes \citep{yoon11,Carlberg:12}.

Given these results, in subsequent sections we restrict our attention to the case of direct encounters of lower mass subhalos with thin streams that are typically generated by the destruction of a globular cluster.

\subsection{Dark matter encounters with thin stellar streams}

\subsubsection{The Dynamics of Gaps in Stellar Streams}

Star stream density variations on scales much smaller than the orbit of the stream are the result of encounters with perturbers (such as dark matter subhalos), the dynamics of the ejection of stars from the progenitor, and compression and expansion of a stream around an orbit.

The response of an infinitely thin stream to an encounter with a relatively low mass perturber is straightforward to calculate analytically
using the impulse approximation, and the results can be generalized to streams with finite width. The results are in good agreement with 
numerical orbit integrations for streams with width to orbital radius ratios of 1:300, which includes the regimes of the two well studied thin streams (Pal 5 and GD-1).  If a section of a stream encounters a massive satellite (e.g., the LMC\index{Large Magellanic Cloud (LMC)}), then that section of the stream will be completely pulled away and spread around the host galaxy.

\begin{figure}[!t]
\begin{center}
\includegraphics[angle=0, scale=0.4]{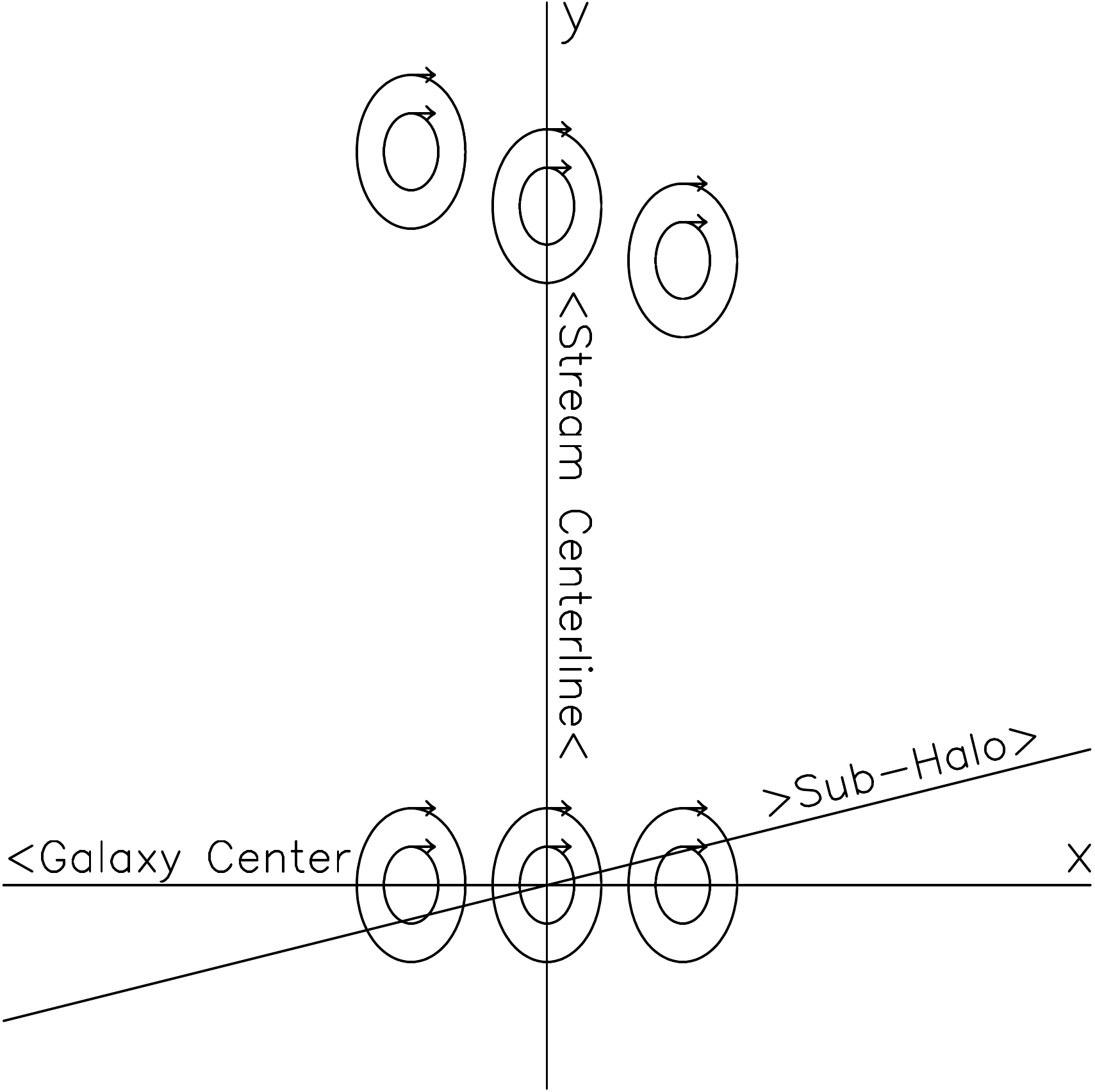}
\end{center}
\caption{A coordinate system to analyze the development of gaps in a stream. The stream
is moving upward along the $y$ axis. The guiding centers of three masses at different locations across the tidal stream are shown at $y=0$.  At the top of the figure, the guiding centers of the same three masses are shown at a later time. The differential rotation of a galaxy means that the guiding centers of particles at larger
radii drift backward. The width of the stream is determined by the combination of epicyclic motions and  spread of guiding centers, the details of which depend on the orbit of the progenitor system \citep{Carlberg:13}.
  In the text, we consider a subhalo that
crosses the stream at $y=0$ at $t=0$. [Reproduced from \citet{Carlberg:13}.] }
\label{fig_coords}
\end{figure}

Figure \ref{fig_coords} introduces a coordinate system for calculating the effects of subhalos passing through a stream. A perturbing mass with a spherically symmetric gravitational potential,  $\Phi(r)$, induces a net velocity change, as a function of distance along the stream of:
\begin{equation}
\Delta\vec{v}(y) = -\int_{-\infty}^{\infty}  \vec{\nabla}\Phi(| \vec{d}(y,t)|)\, dt,
\label{eq_dv}
\end{equation}
where \vec{d}(y,t) is the distance along the stream from the stream crossing point at $(x,y)=(0,0)$.  Equations~\ref{eq_dv} are straightforward to numerically integrate for most radially symmetric density profiles.

To illustrate the behavior, \citet{Carlberg:13} analytically integrated the velocity changes of Equation~\ref{eq_dv} for a perturbing point mass, $M$, moving at speed $(v_x, v_y)$, and crossing a stream that is moving along the $y$ axis at speed $V_y$.
The velocity of the point mass relative to the stream is defined as $v_\parallel=v_y-V_y$ (the velocity parallel to the stream), and
 $v_\perp=v_x$ (toward the stream).
The distance of closest approach of the mass to the stream is the impact parameter, $b$.   
The change in the $v_\parallel$ component of the stream stars produces a change in the relative velocity of the stream and the subhalo given by:,
\begin{equation}
\Delta v_\parallel(y)={\frac {-2{\it GM}{v_\perp}^2 y}
{v \left( v^2 b^2+{v_\perp}^2 y^2 \right)}},
\label{eq_dvpar}
\end{equation}
where $v= \sqrt{v_\parallel^2+v_\perp^2}$ is the speed of the perturbing mass relative to the stream stars.  
This equation has been previously derived in 
\citet{yoon11} 
with slightly different notation.
For the direction of motion perpendicular to the stream, the velocity change is 
$
\Delta v_\perp(y)=-( v_\parallel /v_\perp)
\Delta v_\parallel(y).
$
In the direction of smallest separation between  the perturber and the stream (here called the $z$ direction)  the change is
$
\Delta v_z(y)=(v_\perp^2 y/(v^2b))\Delta v_\parallel(y),
$
which has the same sign at any location along the stream.

\begin{figure}[!t]
\begin{center}
\includegraphics[angle=-90, scale=0.4]{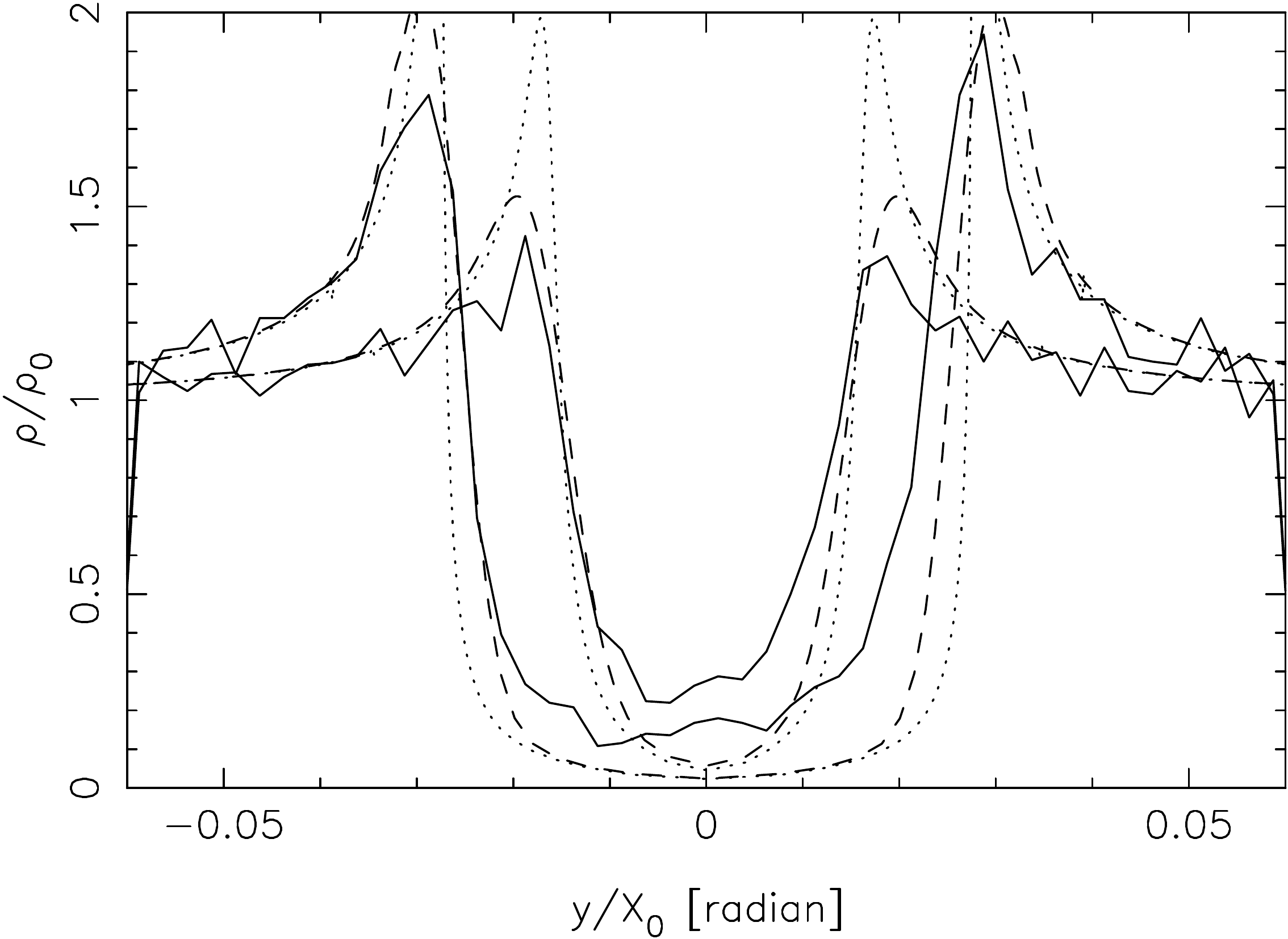}
\end{center}
\caption{The gap at two times (6.4 and 12.8 orbits after the encounter) in a detailed orbit integration (jagged continuous line) in a warm stream of width 0.005 relative to the orbital radius. The dotted line shows the
predicted density profile for a cold stream and the dashed line is the prediction allowing 
for epicyclic motions. The simple theory is in good agreement with the simulation. [Reproduced from \citet{Carlberg:13}.]}
\label{fig_d005}
\end{figure}

The effect of the perturber is to pull particles along the stream towards the crossing point, since $\Delta v_\parallel(y)$ is positive for negative $y$ and negative for positive $y$. In addition, the displacement perpendicular to the stream has
the same dependence on distance along the stream 
and is proportional to 
$-v_\parallel/v_\perp$; that is, the displacement is toward the incoming side below the crossing point (negative $y$) and away above it. 

The velocity change was derived
for a stream moving in a straight line; however, the velocity changes can be applied to a stream moving in a nearly circular orbit. For circular orbits, the particles travel along their guiding centers. The velocity changes along the stream are angular momentum changes which cause stars ahead of the crossing point (positive $y$) to have a reduced angular momentum and hence move to a smaller guiding center which always has a higher rate of angular rotation. Thus the stars ahead of the crossing point are pulled ahead. Similarly, the star behind the crossing point (negative $y$) are moved to lower angular rotation and fall behind. In this way, a gap is formed in the tidal stream.

\begin{figure}[!t]
\begin{center}
\includegraphics[angle=0, scale=0.8]{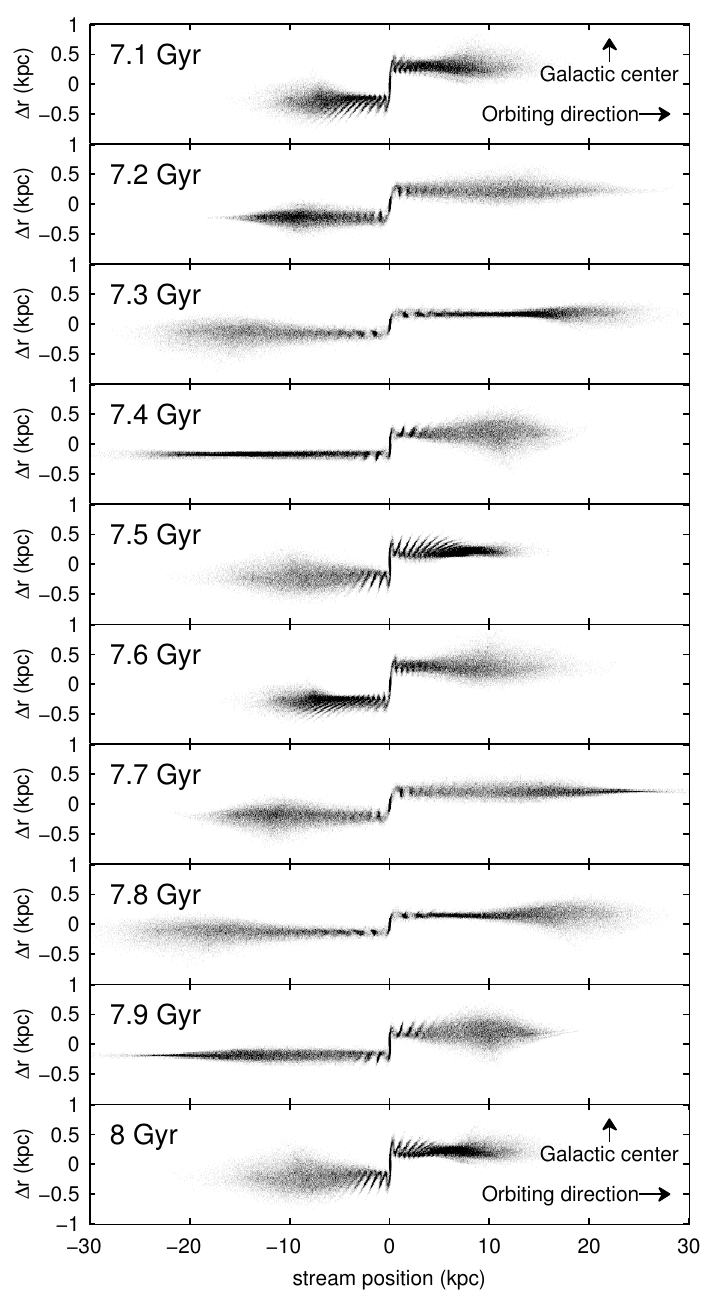}
\includegraphics[angle=0, scale=0.8]{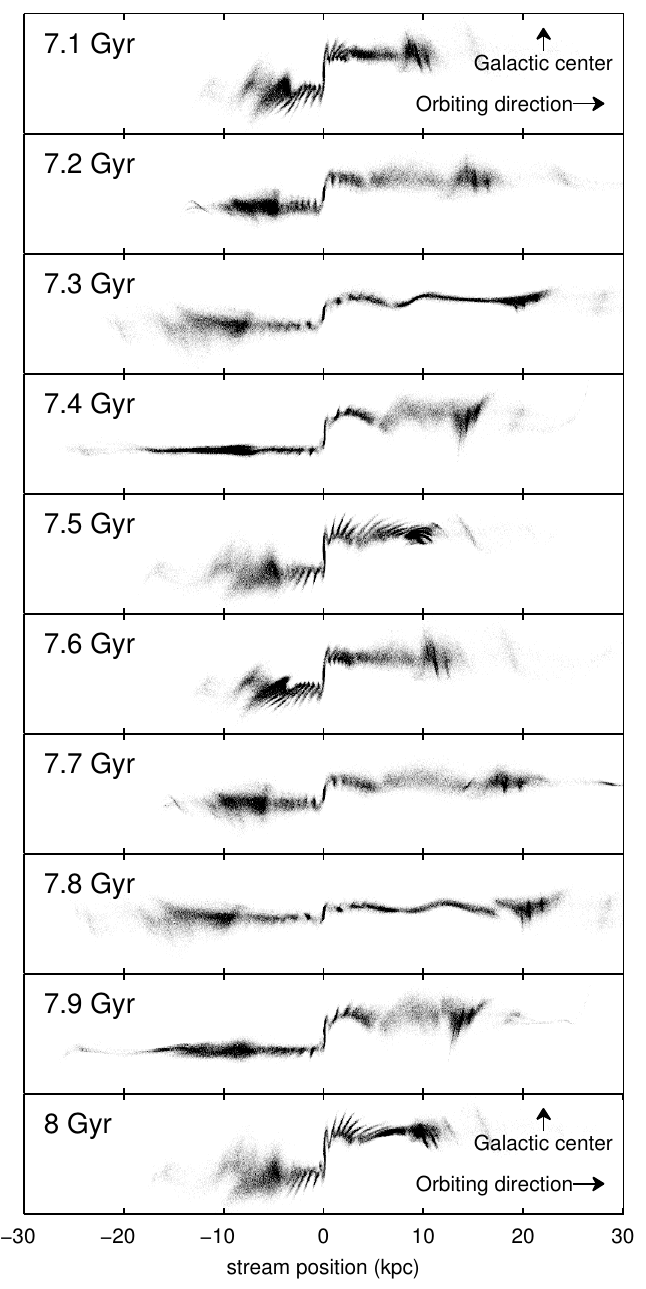}
\caption{Left panels: Simulation of a stream that develops from a globular cluster, projected onto its orbital plane, and transformed to cartesian coordinates. $\Delta r$ is the difference between the distance from the progenitor to the Galactic center and the distance from the stream debris to the Galactic center. Stream position is measured along the tidal stream. Right panels: Same as the left column, but with subhalos included in the simulation, excluding subhalos above $10^8~M_\odot$.}
\label{fig_nbody}
\end{center}
\end{figure}

The equations below are developed for the case of a perturber moving parallel to the orbital plane of the stream, so that the impact parameter is zero. For other orientations of the perturber, the direction of the response of the stream stars changes but the density profile of the resulting gap is essentially the same.

Stars behind the crossing point gain angular momentum and move to a larger guiding center, which has a lower rate of angular rotation, according to:
\begin{equation}
\Omega(y) = {v_c\over X_0}\left[1 + {\Delta v_\parallel(y) \over{v_c}}\right]^{-1}.
\label{eq_Om}
\end{equation}
Hence, the stars begin to move apart, 
and a gap develops over a rotation period. The density profile of the gap can be derived from Equation~\ref{eq_Om}, which gives the change in angular position of the stars with time (see \citealt{Carlberg:13}).
The gap starts with a size comparable to the impact parameter or scale radius of the perturbing object, and 
continues to grow in length with time. 
The material that moves out of the gap piles up on either side and creates a characteristic ``double horned" density profile. 
The infinitely narrow stream can be broadened to a finite width by introducing a Gaussian distribution of epicycles on the stream. Epicycles are often used to describe non-circular orbits as the combination of a guiding center with a circular orbit combined with the motion of an orbiting body on an ``epicycle" orbit around the guiding center. In this case, the epicycles are introduced to give the stream width; the particles will wiggle back and forth around the center of the tidal stream.  The cold stream density profile of a gap is then convolved with the appropriate Gaussian distribution along the stream. Numerical integrations show good agreement with this simple theory (Figure~\ref{fig_d005}). 
Recently Erkal \& Belokurov (2014) have extended this analysis into the mildly nonlinear regime, demonstrating that the
folding of the stream leads to caustics in the density profile and that the growth in the width of the stream with time slows from $t$ to $\sqrt{t}$.

\begin{figure}[!t]
\centering
\includegraphics[angle=0, width=1.0\textwidth]{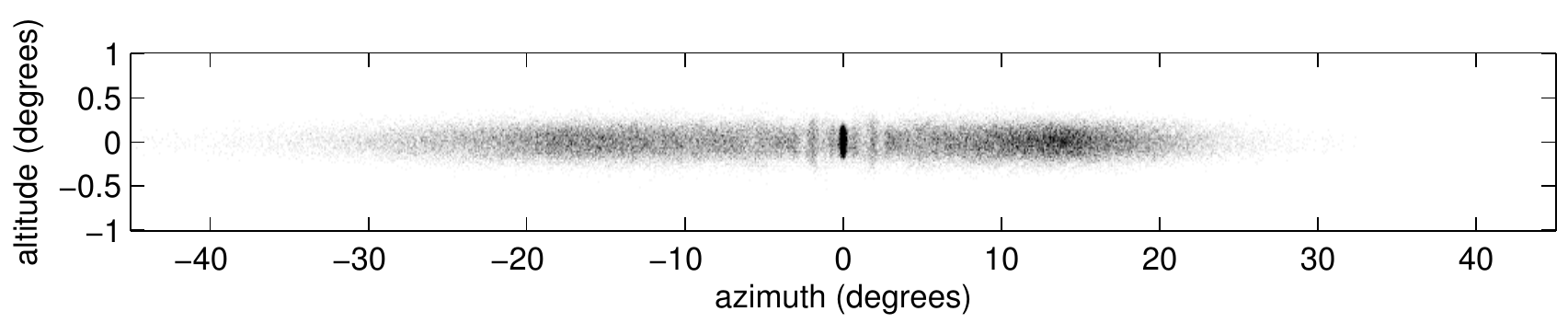}
\includegraphics[angle=0, width=1.0\textwidth]{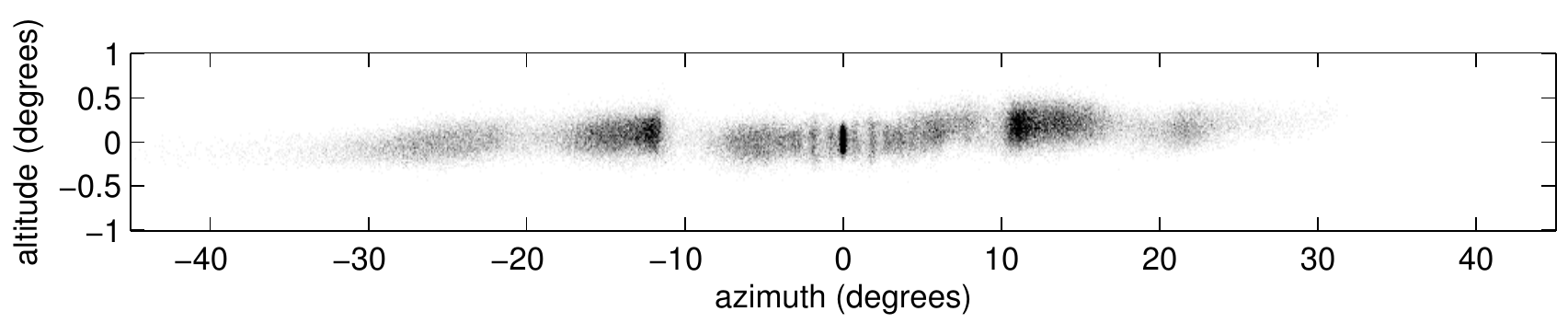}
\caption{The view from the center of the galaxy of the 7.5 Gyr timeslice of Figure~\ref{fig_nbody}. 
The top panel shows the simulated tidal stream without subhalos, and the bottom panel includes a cosmologically motivated set of subhalos in the simulation.
This projection is likely closer to a typical view of a stream outside the solar circle. The youngest part of the stream,
close to the progenitor, contains epicyclic oscillations which phase mix away. The stream becomes older further away and is dominated by subhalo induced gaps.}
\label{fig_GCnbody}
\end{figure}

More general simulations of the dissolution of a progenitor system in a realistic halo potential are essential to give a more realistic view of a stream. Because they contain no unseen dark matter to confuse the dynamical situation, disrupting globular clusters are the most straightforward systems to model.
Nevertheless, even this fairly well defined situation is yet to be completely understood. This is 
largely a result of the range of orbits and potentials that need to be investigated as well as the 
details of how the progenitor is modeled.
Kupper and collaborators have shown that when stars leave a globular cluster they stream through 
the Lagrange points with a fairly narrow spread of (primarily radial) velocities.
Consequently the stars follow a cycloidal path. However, the dominant effect for clusters on non-circular orbits is that the mass loss varies around the orbit, leading to ``spurs'' which oscillate above 
and below the centerline of the stream at the epicyclic frequency. 

One important outcome of a number of simulations is that the range of angular momentum within the streams from a globular cluster is  quite small. This implies that there is an almost unique association between time since ejection from the globular cluster into the stream, and distance from the progenitor along the stream. Furthermore, there is very little shear present at any given distance along the stream, meaning that features in the stream are not blurred out significantly with time \citep{bovy14, Carlberg:14}.

Although in principle it is possible to work out the distribution of the number of gaps of a given size in a stream, the number of effects that need to be modeled mean that it is easier to do either a partially numerical integration, or a complete simulation. 
The basic idea is straightforward: small sub-halos cause small gaps (which subsequently grow with time).  Therefore we expect that there will be a spectrum of gap sizes rising as a power law towards smaller gaps until the spectrum rolls over because random motions in the stream blur out gaps that are smaller than about the stream width.  \citet{CG:13} presented a semi-analytic calculation of this spectrum.
Two important effects are left out of the calculation.  First, the gaps will overlap, meaning that the semi-analytic calculation is an upper limit. 
Second, some of the gaps might not be observable given a finite number of detected stars in the stream.

\begin{figure}[!t]
\begin{center}
\includegraphics[angle=-90, scale=0.4]{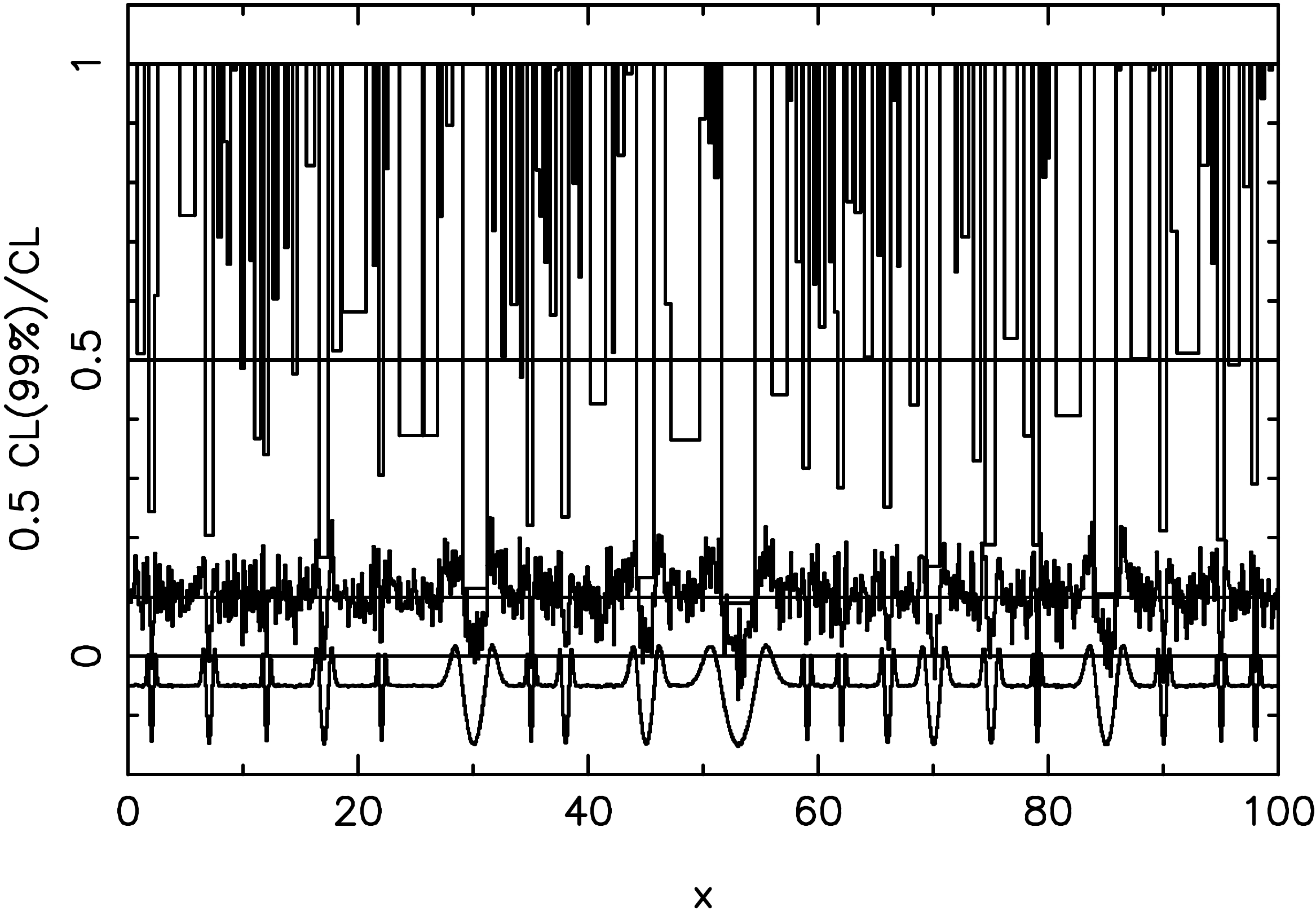}
\end{center}
\caption{A demonstration of the gap filter applied to artificial data with noise characteristics comparable to GD-1. The filter systematically over-estimates the number of gaps by about 44\%. The filter is a well-defined approach to finding and measuring gaps, but this version is subject to (calibratable) systematic errors.}
\label{fig_test}
\end{figure}

\subsubsection{The cumulative effect of sub-halos}

\citet{CG:13} refined the cold-stream analysis of \citet{Carlberg:12} to
predict that the number of gaps created per unit time per unit length in the stream (the gap creation rate, $R_\cup$) as a function of galactocentric distance, $r$, in units of 30 kpc, and 
$M_8=M/10^8 M_\odot$ of,
\begin{equation}
R_\cup =0.00433r^{0.26} M_8^{-0.36} {\rm kpc}^{-1} {\rm Gyr}^{-1}.
\label{eq_R}
\end{equation}
At an encounter age of around 4 Gyr, sub-halos  of mass $M_8$ create gaps of mean length $\ell$,
\begin{equation}
\ell = 9.57 r^{0.16} M_8^{0.31} {\rm kpc}.
\label{eq_l}
\end{equation}
Eliminating $M_8$ between Equations~\ref{eq_R} and \ref{eq_l}, we
find the gap creation rate as a function of gap size is:
\begin{equation}
{dn(\ell)\over{dt}} d\ell= 0.060 r^{0.44} \ell^{-1.16} \, {\rm kpc}^{-1} {\rm Gyr}^{-1} {d\ell\over \ell}
\label{eq_rate}
\end{equation}
for gaps of size $\ell$, which is measured in kpc and the variable $r$ is scaled to 30 kpc. If we evaluate this at 
15 kpc for a stream of 4 Gyr age and integrate over all gap sizes larger than $\ell$ we find the 
fraction
of the stream length that has gaps is
\begin{equation}
f(>\ell) = 1.1  \ell^{-0.16} {T\over {4~{\rm Gyr}}}.
\label{eq_cum}
\end{equation}
That is, for streams within about 30~kpc of the Galactic center, every position along a star stream of this age has been 
affected by a subhalo \citep{Carlberg:09}.
The dependence on gap size is very weak. 
The low mass subhalos heat the stream and to some degree frustrate and
complicate the formation of subsequent small gaps.  However, the stream remains intact.
The larger subhalos that create
gaps of several kpc in size cause sufficiently large perturbations perpendicular to the
stream that
it becomes possible for the orbits of stream segments to diverge,
 particularly if the overall potential is strongly triaxial (or more complicated) and/or time evolving.

Work has begun on the dynamical modeling of streams in realistic cosmological halos.
\citet{cooper10} examined streams already formed within a cosmological simulation
which usefully illustrate the complicated time evolution of the stream shape.
However, these simulations do not have the mass resolution to follow cool streams or
globular cluster dissolution. \citet{bonaca14} have published a realistic, but approximate, approach
to following a dissolving cluster in an evolving halo.  
One basic outcome is that streams that orbit in the outer parts of the galaxy halo are 
much less disturbed than those that orbit within the much more dynamic inner parts of the galaxy halo, as expected. 
The buildup of the visible stellar mass which dominates the potential field of the 
inner galaxy has yet to be modeled.

\subsubsection{Detecting Gaps in Stellar Streams}

The calculations of gap shapes above usefully predict that gaps should have a double-horned profile with an integral along their length of zero, 
meaning the mass displaced from the gap is simply piled up on either side of the stream.
Gap finding then consists of running filters of all widths along the stream to find regions where there is a good match, identified as local peaks in the filtered distribution. 
To quantify the statistical confidence, the same filter is run through a density distribution with the same noise properties.
 The resulting distribution of peaks is sorted to identify what level constitutes 99\% confidence that a peak is not a false positive.
Although this filtering procedure gives gap-finding a statistical foundation, the current procedures could be improved. 
In particular, the rate of false positives is currently about 30\% of the peaks. This factor can be included as a correction, but reducing their number would be helpful.
 The shape of the gap filter is currently essentially an informed approximation and is not driven by the characteristics of the gaps in the data.  
That is, there is no current empirical approach to generate a gap spread function, comparable to the point spread function of a star that can be empirically determined in image data.  
Work is now beginning to undertake more extensive simulations; placing those results into a simulated sky will greatly improve the understanding of gap finding techniques.

\begin{figure}[!t]
\begin{center}
\includegraphics[angle=0, scale=0.4]{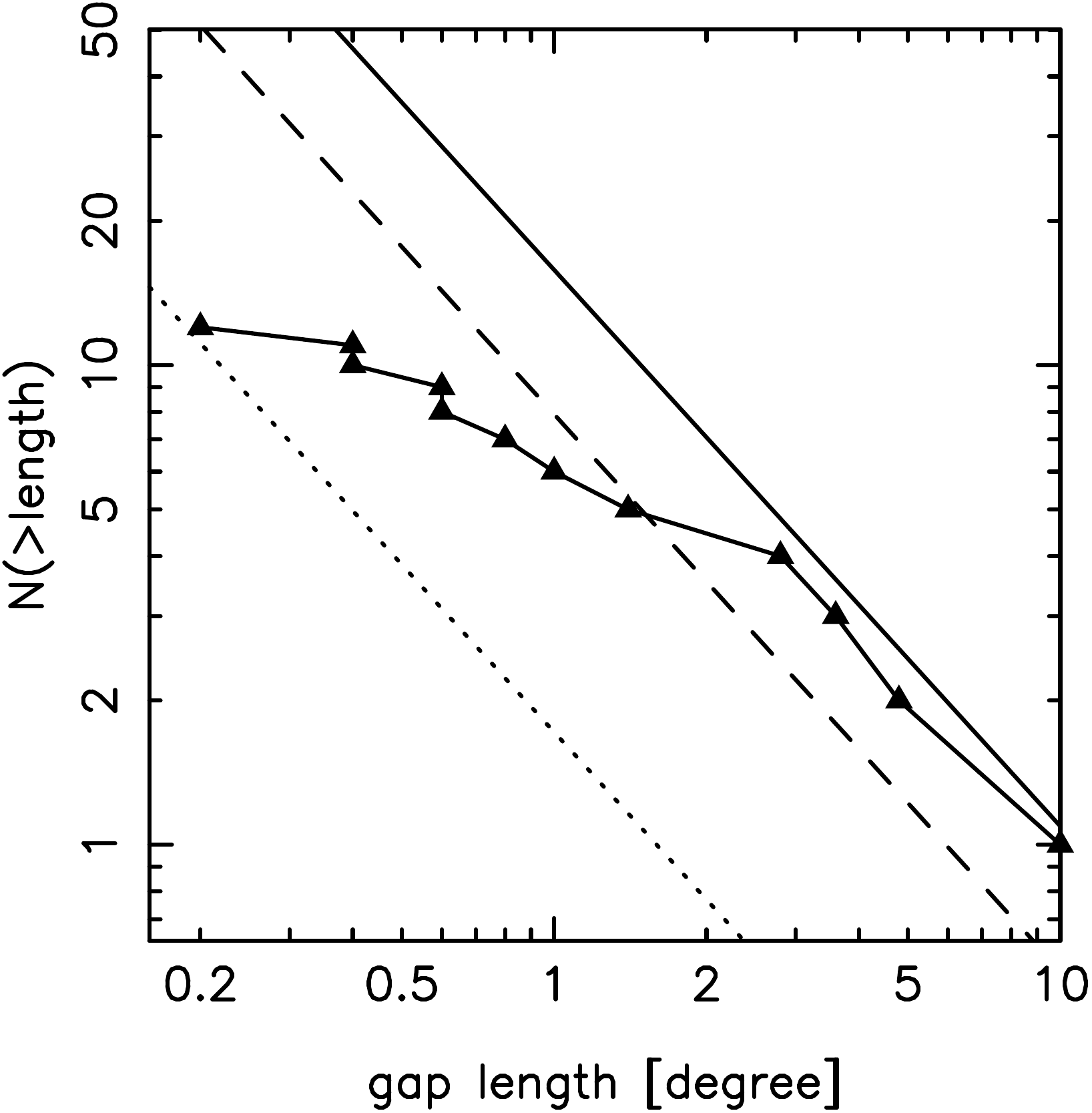}
\end{center}
\caption{A comparison of the recovered GD-1 cumulative distribution of gap sizes with the semi-analytic estimate of the expected numbers for various assumed mean stream ages.}
\label{fig_steerable}
\end{figure}

\subsection{Current observations and future prospects}

Currently there are only two really well studied thin streams: the Pal~5 stream, emanating  from the tidal lobes of the Pal~5 globular cluster, and the GD-1 stream, having no known progenitor. 
 
There are three primary reasons for density variations in streams: the dynamics of mass loss from the progenitor, variations in velocity around the stream's orbit, and perturbations caused by encounters of the stream with any massive object. Massive objects range from baryonic structures to dark matter sub-halos, possibly containing visible stars or gas. 
The orbital effects are necessarily smooth variations around the orbit and hence on scales much larger than those which vary with the comparable size scales of the progenitors and the dark matter sub-halos of interest, which we review below.

So far streams are detected through measurements of sky density, most often in the SDSS survey where the requirement for photometric colors precise to about 10\% or better limits the data to about 21-22 magnitude in the SDSS system.
 Reaching large numbers of stars requires getting to at least the bottom of the red giant branch and ideally to main sequence turn off stars (or beyond) at absolute magnitudes of +5 or so. 
The outcome is that streams tend to be found 
at distances in the range of 10-20~kpc in the currently available data.
Even with the very good optimal weighting procedures of
\citet{rockosi02},
which dramatically reduce the weight of background stars, 
the summed weight of stars in the stream is typically about 10-30\% of the backgrounds.  To obtain a local signal-to-noise of about one usually requires binning over the entire width of narrow streams, so no information on 2D structure is available with current data. As deeper images, with more filters and eventually kinematic data become available, the local signal to noise will dramatically rise. Consequently, the currently available data is usually restricted to being a one dimensional density along the path of the stream. 

Even a 60$^\circ$\ long thin stream like GD-1 is expected to have only about a dozen or so detectable gaps over its visible length.
Detecting more streams, which means going to larger Galactic radii, is a key element of the future of the field. In addition, more filter bands will allow improved optimal photometric matched filtering to include metallicity information to further suppress the foreground and background stars of the same temperature and luminosity. And finally, as better kinematic data slowly become available (note that Gaia will only reach about 20th magnitude, which is the regime where streams 
discovered in the SDSS pick up much of their signal), 
the use of improved distances and velocities for each individual star will allow us to more accurately identify which stars are in the stream, and will thus improve the detail with which models can be matched to the data. Moreover, it will then be possible to use kinematic signatures of gaps (a sideways S in velocity space)  
to find gaps and characterize the perturbers that caused them.

There is a vast array of planned all-sky imaging and spectroscopic surveys from both the ground and space that will transform our knowledge of stellar streams over the next decade. First, we will find new streams in the southern hemisphere, links with known streams in the north, and, in both hemispheres, find streams out to about 100 kpc with higher signal-to-noise than current data. 
Spectra will provide astrophysical information about the nature of the stream progenitors and stream kinematics, which will be particularly powerful in combination with proper motion and distance data. The challenge will then be to make use of these data, which will be somewhat noisy by theoretical standards, to put new and interesting constraints on the nature of both the smooth large scale potential of the galaxy, and the small scale variations in the potential expected in a $\Lambda$CDM universe.

\begin{acknowledgement}
KVJ thanks her postdocs and graduate students for invaluable discussions throughout the year (Andreas Kuepper, Allyson Sheffield, Lauren Corlies, Adrian Price-Whelan, David Hendel and Sarah Pearson).
Her work on this volume was supported in part by NSF grant AST-1312196.
RGC thanks his graduate student Wayne Ngan and support from CIfAR and NSERC is gratefully acknowledged.
\end{acknowledgement}

%
%
%


\end{document}